\documentclass{article}%
\usepackage{amsfonts}
\usepackage{amsmath}
\usepackage{amssymb}
\usepackage{graphicx}%
\providecommand{\U}[1]{\protect\rule{.1in}{.1in}}

\begin{document}

\title{Remarks on Quantum Aspects of 3D-Gravity in the First-Order Formalism}
\author{L. M. de Moraes\thanks{moraes@cbpf.br}, J. A.
Helay\"{e}l-Neto\thanks{helayel@cbpf.br}, V. J. V\'{a}squez Otoya
\thanks{vjose@cbpf.br}.\\Centro Brasileiro de Pesquisas F\'{\i}sicas (CBPF), R. Dr Xavier Sigaud, 150\\Urca, Rio de Janeiro, RJ, Brazil - 22290-180}
\maketitle

\begin{abstract}
In this paper, we reassess the issue of working out the propagators and
identifying the spectrum of excitations associated to the vielbein and spin
connection of (1+2)-D gravity in the presence of torsion by adopting the
first-order formulation. A number of peculiarities is pointed out whenever the
Chern-Simons term is taken into account along with the possible bilinear terms
in the torsion tensor. We present a procedure to derive the full set of
propagators, based on a set of spin-type operators, and we discuss under which
conditions the pole of these tree-level 2-point functions correspond to
physical excitations.

\end{abstract}

\section{Introduction}

The need to better understand gauge fields has lead to an widespread use of
local transformations due the natural manner gauge fields appear in it. In the
attempt to write (1+2)-D gravity as a gauge theory, the formulation requires
some specific technicalities, by virtue of the possibility of including the
so-called (topological) Chern-Simons term. Adopting the Poincar\'{e} group as
the local gauge group, one naturally obtains the curvature and torsion tensor
by means of the Cartan%
\'{}%
s structure equations. The translational part of the Poincar\'{e} group is
represented by the vielbein gauge fields, $e_{\alpha}{}^{\mathsf{a}}$, which
are also diffeomorphic invariant under general coordinate transformations, and
the Lorentzian part --- realizing the equivalence principle --- given by the
spin connection gauge fields, $\omega_{\alpha}{}^{\mathsf{ab}}$. The vielbein
fields associate to each point a locally flat coordinate system and the spin
connection relates any two local Lorentz coordinate systems at the given point.

This formalism is believed to be completely equivalent to the formalism that
employs affine connections and define curvature and torsion by means of it.
There is a great deal of results that confirm this, mainly at the level of
expressions to the curvature and torsion. At the classical level, this
equivalence is indeed true. However, where we go over to the quantum
field-theoretic version, there appear remarkable differences and we must
indeed adopt the vielbein and spin connection as the independent fundamental
degrees of freedom \cite{two}.

In order to investigate this further, we begin with the analysis of a
traditional (1+2)-D gravity model previously done by two of us \cite{old
article}, where we studied the inclusion of torsion in three-dimensional
Einstein-Chern-Simons gravity and added up higher-derivative terms, all in the
affine connection formalism. In this work, due to invertibility problems that
appear in the theory using the local formalism, we are forced to change to a
simpler Lagrangian, where we consider the Einstein-Chern-Simons Lagrangian
with torsion algebraic terms only.

Due to the importance of the torsion terms, it is worthwhile to remember that
torsion was introduced by E. Cartan in 1922, as the antisymmetric part of the
affine connection and was recognized by him as a geometric object related to
an intrinsic angular moment of matter. After the introduction of the spin
concept, it was suggested that torsion should mediate a contact interaction
between spinning particles without propagation in matter-free space
\cite{hehl1},\cite{hehl2}. Later,\ due the fact that at the microscopic level
particles are classified by their mass and spin according to the Poincar\'{e}
group, gauge theories of General Relativity were developed that brings in it
dynamic torsion \cite{novello},\cite{sabbata2}. These theories are motivated
by the requirement that the Dirac equation in a gravitational field preserves
local invariance under Lorentz transformations which yields, across the
minimal coupling, a direct interaction between torsion and fermions.
Observational constraints for a propagating torsion and its matter
interactions are discussed in \cite{carrol},\cite{hammond},\cite{shapiro}%
,\cite{sabbata}.

Our work is organised according to the following outline: in Section 2, we
present a quick review of the Einstein-Cartan formalism, with the purpose of
fixing notation and setting our conventions. Next, the general model and the
decomposition of the action in terms of spin operators is the subject of
Section 3, where we point out a serious problem related to a spin-2
excitation. This motivates us to introduce and to analyse a number of torsion
terms in the action, which is done in Section 4. In Section 5, we come to the
task of computing the propagators and we analyse thereby their poles with the
corresponding residues, in order to locate the physically relevant regions in
the parameter space. Finally, in Section 6, we present our Concluding
Comments, with a critical discussion on our main results and possible issues
for future investigation.

\section{Well-known Results on the Einstein-Cartan Approach}

A Riemann-Cartan space-time \cite{hehl1},\cite{sabbata}\ is defined as a
manifold where the covariant derivative of the metric field exists and is
given by:%

\begin{equation}
\nabla_{\gamma}g_{\alpha\beta}(x)=0, \label{um}%
\end{equation}
where this equation defines the so called metric-compatible affine connection,
$\Gamma_{\alpha\beta}{}^{\gamma}$; it allows the presence of torsion, given by
the antisymmetric part of the affine connection,%

\begin{equation}
\mathcal{T}_{\alpha\beta}{}^{\gamma}=2\Gamma_{\lbrack\alpha\beta]}{}^{\gamma}.
\label{dois}%
\end{equation}

We then have:%
\begin{equation}
\Gamma_{\alpha\beta}{}^{\gamma}=\left\{  _{\alpha\beta}^{\ \gamma}\right\}
+K_{\alpha\beta}{}^{\gamma}, \label{tres}%
\end{equation}
where $\left\{  _{\alpha\beta}^{\ \gamma}\right\}  $ is the Christoffel
symbol, which is completely determined by the metric,%

\begin{equation}
\left\{  _{\alpha\beta}^{\ \gamma}\right\}  =\frac{1}{2}g^{\gamma\lambda
}(\partial_{\alpha}g_{\lambda\beta}+\partial_{\beta}g_{\alpha\lambda}%
-\partial_{\lambda}g_{\alpha\beta})
\end{equation}
and%

\begin{equation}
K_{\alpha\beta}{}^{\gamma}=\frac{1}{2}(\mathcal{T}_{\alpha\beta}{}^{\gamma
}+\mathcal{T}^{\ \gamma}{}_{\alpha\beta}-\mathcal{T}_{\beta}{}^{\gamma}%
{}_{\alpha})
\end{equation}
is the contortion tensor, antisymmetric in the last two indices.

In order to study local properties one introduces (in our specific (1+2)-D
case) the dreinbein vector fields, $e_{\alpha}{}^{\mathsf{a}}(x)$,\ that spans
at any given point the local Minkowski space-time, which in this work has
metric: $\eta_{\mathsf{ab}}=diag(1,-1,-1)$.

The introduction of the tangent Minkowski space-time allows local Lorentz
transformations on geometrical objects (with Latin index). In order to render
these objects invariant under local Lorentz rotations, one introduces the spin
connection $\omega_{\mathsf{\gamma b}}{}^{\mathsf{c}}$. The covariant
derivative of the dreinbein then reads:%

\begin{equation}
\nabla_{\gamma}e_{\alpha}{}^{\mathsf{a}}=D_{\gamma}e_{\alpha}{}^{\mathsf{a}%
}-\Gamma_{\gamma\alpha}{}^{\lambda}e_{\lambda}{}^{\mathsf{a}}=0, \label{seis}%
\end{equation}
where $D_{\gamma}e_{\alpha}{}^{\mathsf{a}}=\partial_{\gamma}e_{\alpha}%
{}^{\mathsf{a}}+\omega_{\gamma\mathsf{i}}{}^{\mathsf{a}}e_{\alpha}%
{}^{\mathsf{i}}$ is the Lorentz covariant derivative.

One finds from eq.(\ref{seis}) that the affine connection can then be written as:%

\begin{equation}
\Gamma_{\alpha\beta}{}^{\gamma}=e_{\mathsf{j}}{}^{\gamma}D_{\alpha}e_{\beta}%
{}^{\mathsf{j}}, \label{sete}%
\end{equation}
and the torsion tensor, eq.(\ref{dois}), reads%

\begin{equation}
\mathcal{T}_{\alpha\beta}{}^{\gamma}=2\Gamma_{\lbrack\alpha\beta]}{}^{\gamma
}=e_{\mathsf{j}}{}^{\gamma}(\partial_{\alpha}e_{\beta}{}^{\mathsf{j}}%
-\partial_{\beta}e_{\alpha}{}^{\mathsf{j}}+\omega_{\alpha\mathsf{i}}%
{}^{\mathsf{j}}e_{\beta}{}^{\mathsf{i}}-\omega_{\beta\mathsf{i}}{}%
^{\mathsf{j}}e_{\alpha}{}^{\mathsf{i}}). \label{doze}%
\end{equation}

As known, the curvature tensors and scalar are given in terms of the affine
connection by the expressions:%

\begin{equation}
\mathcal{R}_{\mu\alpha\beta}{}^{\nu}=\partial_{\mu}\Gamma_{\alpha\beta}{}%
^{\nu}-\partial_{\alpha}\Gamma_{\mu\beta}{}^{\nu}+\Gamma_{\mu\rho}{}^{\nu
}\Gamma_{\alpha\beta}{}^{\rho}-\Gamma_{\alpha\rho}{}^{\nu}\Gamma_{\mu\beta}%
{}^{\rho},
\end{equation}

\begin{equation}
\mathcal{R}_{\alpha\beta}=\mathcal{R}_{\mu\alpha\beta}{}^{\mu}=\partial_{\mu
}\Gamma_{\alpha\beta}{}^{\mu}-\partial_{\alpha}\Gamma_{\mu\beta}{}^{\mu
}+\Gamma_{\mu\rho}{}^{\mu}\Gamma_{\alpha\beta}{}^{\rho}-\Gamma_{\alpha\rho}%
{}^{\mu}\Gamma_{\mu\beta}{}^{\rho}%
\end{equation}
and%

\begin{equation}
\mathcal{R}=g^{\alpha\beta}\mathcal{R}_{\alpha\beta}.
\end{equation}

In terms of the spin connection,%

\begin{equation}
\mathcal{R}_{\mu\alpha\beta}{}^{\nu}=e_{\beta}{}^{\mathsf{i}}e_{\mathsf{j}}%
{}^{\nu}(\partial_{\mu}\omega_{\alpha\mathsf{i}}{}^{\mathsf{j}}-\partial
_{\alpha}\omega_{\mu\mathsf{i}}{}^{\mathsf{j}}+\omega_{\mu\mathsf{k}}%
{}^{\mathsf{j}}\omega_{\alpha\mathsf{i}}{}^{\mathsf{k}}-\omega_{\alpha
\mathsf{k}}{}^{\mathsf{j}}\omega_{\mu\mathsf{i}}{}^{\mathsf{k}}),
\end{equation}%
\begin{equation}
\mathcal{R}_{\alpha\beta}=e_{\beta}{}^{\mathsf{i}}e_{\mathsf{j}}{}^{\mu
}(\partial_{\mu}\omega_{\alpha\mathsf{i}}{}^{\mathsf{j}}-\partial_{\alpha
}\omega_{\mu\mathsf{i}}{}^{\mathsf{j}}+\omega_{\mu\mathsf{k}}{}^{\mathsf{j}%
}\omega_{\alpha\mathsf{i}}{}^{\mathsf{k}}-\omega_{\alpha\mathsf{k}}%
{}^{\mathsf{j}}\omega_{\mu\mathsf{i}}{}^{\mathsf{k}}) \label{oito}%
\end{equation}
and%

\begin{equation}
\mathcal{R}=\eta^{\mathsf{ai}}e_{\mathsf{a}}{}^{\alpha}e_{\mathsf{j}}{}^{\mu
}(\partial_{\mu}\omega_{\alpha\mathsf{i}}{}^{\mathsf{j}}-\partial_{\alpha
}\omega_{\mu\mathsf{i}}{}^{\mathsf{j}}+\omega_{\mu\mathsf{k}}{}^{\mathsf{j}%
}\omega_{\alpha\mathsf{i}}{}^{\mathsf{k}}-\omega_{\alpha\mathsf{k}}%
{}^{\mathsf{j}}\omega_{\mu\mathsf{i}}{}^{\mathsf{k}}). \label{nove}%
\end{equation}

\section{A Problem Related to a Spin-2 Excitation}

We start off with the three-dimensional action for topologically massive gravity:%

\begin{equation}
\mathcal{S=}\int d^{3}x\ e\left(  \mathsf{a}_{1}\mathcal{R+}\mathsf{a}%
_{2}\mathcal{R}^{2}+\mathsf{a}_{3}\mathcal{R}_{\alpha\beta}\mathcal{R}%
^{\alpha\beta}+\mathsf{a}_{4}\mathcal{L}_{CS}\right)  , \label{dez}%
\end{equation}
where%

\begin{equation}
\mathcal{L}_{CS}=\varepsilon^{\alpha\beta\gamma}\Gamma_{\gamma\delta}%
{}^{\lambda}\left(  \partial_{\alpha}\Gamma_{\lambda\beta}{}^{\delta}+\frac
{2}{3}\Gamma_{\alpha\rho}{}^{\delta}\Gamma_{\beta\lambda}{}^{\rho}\right)  ,
\label{quatorze}%
\end{equation}
is the topological Chern-Simons term and%

\begin{equation}
\varepsilon^{\alpha\beta\gamma}=\frac{\epsilon^{\alpha\beta\gamma}}{e}%
\end{equation}
is the completely antisymmetric tensor in (1+2)-D, with $\epsilon^{\alpha
\beta\gamma}$ the Levi-Civita tensor density in the flat space and $e=\sqrt
{g}$ where $g=\det(g_{\alpha\beta})=\eta e^{2}$. $\mathsf{a}_{1}$,
$\mathsf{a}_{2}$ and $\mathsf{a}_{3}$ are free coefficients, whereas
$\mathsf{a}_{4}$ is the Chern-Simons parameter.

For a beauty discussion of theories with Chern-Simons term see \cite{zanelli}.

As the Riemann tensor, $\mathcal{R}_{\mu\alpha\beta}{}^{\nu}$, has the same
number of independent \ components as the Ricci tensor, $\mathcal{R}%
_{\alpha\beta}$, in three dimensions, a term squared in $\mathcal{R}%
_{\mu\alpha\beta}{}^{\nu}$ is not necessary in the action.

In \cite{old article}, we wrote the affine connection as in eq.(\ref{tres}),
further decomposing the torsion in its SO(1,2) irreducible components: a
scalar from the totally antisymmetric part, a three-vector from the trace and
a symmetric traceless rank-2 tensor. With this procedure, we have obtained a
particle spectrum where only massive excitations of spin-2 associated with the
linearized gravitational field $h^{\alpha\beta}$ and with the symmetric part
of the torsion field had dynamics that preserved the unitarity of the theory
for some values of the action parameters.

In this section, we reconsider the action (\ref{dez}) but, contrary to what we
have done in \cite{old article}, we propose to adopt in the first-order
formulation, dropping the torsion as our fundamental excitation and electing
the dreinbein and the spin connection as the fundamental quantum fields.

Now, making use of equations (\ref{sete}),(\ref{oito}) and (\ref{nove}), along
with the weak field approximation to the gravitational field,%

\begin{equation}
e_{\alpha}{}^{\mathsf{a}}=\delta_{\alpha}{}^{\mathsf{a}}+\frac{k}{2}h_{\alpha
}{}^{\mathsf{a}}\left(  \Rightarrow g_{\alpha\beta}=\eta_{\alpha\beta
}+kh_{a\beta}\right)  ,
\end{equation}
and the spin connection decomposition,%

\begin{equation}
\omega_{\mathsf{a}}{}^{\mathsf{bc}}=\epsilon^{\mathsf{bcd}}Y_{\mathsf{ad}},
\label{quinze}%
\end{equation}
which can be further split in,%

\begin{equation}
Y_{\mathsf{ab}}=y_{\mathsf{ab}}+\mathcal{Y}_{\mathsf{ab}}\ ;\ y_{\mathsf{ab}%
}=Y_{(\mathsf{ab)}}\ \ ,\ \ \mathcal{Y}_{\mathsf{ab}}=Y_{[\mathsf{ab]}}
\label{dezesseis}%
\end{equation}
and%

\begin{equation}
\mathcal{Y}_{\mathsf{ab}}=\epsilon_{\mathsf{abc}}y^{\mathsf{c}}\Rightarrow
y_{\mathsf{a}}=\frac{1}{2}\epsilon_{\mathsf{abc}}\mathcal{Y}^{\mathsf{bc}},
\label{dezessete}%
\end{equation}
we can write the action (\ref{dez}), to which the gauge-fixing terms have been added,%

\begin{equation}
\mathcal{L}_{GF-diff}=\lambda F_{\mathsf{a}}F^{\mathsf{a}}\ ,\ F_{\mathsf{a}%
}=\partial_{\mathsf{b}}\left(  h_{\mathsf{a}}^{\mathsf{b}}-\frac{1}{2}%
\delta_{\mathsf{a}}^{\mathsf{b}}h_{\mathsf{c}}^{\mathsf{c}}\right)  ,
\end{equation}
and%

\begin{equation}
\mathcal{L}_{GF-LL}=\xi\left(  \partial^{\mu}\omega_{\mu}{}^{\mathsf{ab}%
}\partial^{\nu}\omega_{\nu\mathsf{ab}}\right)  ,
\end{equation}

in the convenient linearized form%

\begin{equation}
\mathcal{S}\mathcal{=}\int d^{3}x\frac{1}{2}\Phi^{T}M\Phi\ ,\ \Phi=\left(
\begin{array}
[c]{c}%
y^{\mathsf{cd}}\\
y^{\mathsf{c}}\\
h^{\mathsf{cd}}%
\end{array}
\right)  .
\end{equation}

The wave operator $M$, being expressed in an extension of the spin-projection
operator formalism introduced in \cite{rivers},\cite{nieuwenhuizen},\cite{old
article}. Five additional operators coming from the $y^{\mathsf{a}}$ and
Chern-Simons terms are needed. The six operator for a rank-2 symmetric tensor
in 3D are given by:%

\[
\mathsf{P}_{\mathsf{ab,cd}}^{(2)}=\frac{1}{2}(\theta_{\mathsf{ac}}%
\theta_{\mathsf{bd}}+\theta_{\mathsf{ad}}\theta_{\mathsf{bc}})-\frac{1}%
{2}\theta_{\mathsf{ab}}\theta_{\mathsf{cd}},
\]

\[
\mathsf{P}_{\mathsf{ab,cd}}^{(1-m)}=\frac{1}{2}(\theta_{\mathsf{ac}}%
\omega_{\mathsf{bd}}+\theta_{\mathsf{ad}}\omega_{\mathsf{bc}}+\theta
_{\mathsf{bc}}\omega_{\mathsf{ad}}+\theta_{\mathsf{bc}}\omega_{\mathsf{ad}}),
\]

\begin{equation}
\mathsf{P}_{\mathsf{ab,cd}}^{(0-s)}=\frac{1}{2}\theta_{\mathsf{ab}}%
\theta_{\mathsf{cd}},
\end{equation}

\[
\mathsf{P}_{\mathsf{ab,cd}}^{(0-w)}=\omega_{\mathsf{ab}}\omega_{\mathsf{cd}},
\]

\[
\mathsf{P}_{\mathsf{ab,cd}}^{(0-sw)}=\frac{1}{\sqrt{2}}\theta_{\mathsf{ab}%
}\omega_{\mathsf{cd}}%
\]
and%

\[
\mathsf{P}_{\mathsf{ab,cd}}^{(0-ws)}=\frac{1}{\sqrt{2}}\omega_{\mathsf{ab}%
}\theta_{\mathsf{cd}},
\]
where $\theta_{\mathsf{ab}}$ is the transverse and $\omega_{\mathsf{ab}}$ is
the longitudinal projector operators for vectors. The others five operators are:%

\[
\mathsf{S}_{\mathsf{ab,cd}}^{(2a)}=(\epsilon_{\mathsf{ace}}\theta
_{\mathsf{bd}}+\epsilon_{\mathsf{ade}}\theta_{\mathsf{bc}}+\epsilon
_{\mathsf{bce}}\theta_{\mathsf{ad}}+\epsilon_{\mathsf{bce}}\theta
_{\mathsf{ad}})\partial^{\mathsf{e}},
\]

\[
\mathsf{R}_{\mathsf{ab,cd}}^{(1a)}=(\epsilon_{\mathsf{ace}}\omega
_{\mathsf{bd}}+\epsilon_{\mathsf{ade}}\omega_{\mathsf{bc}}+\epsilon
_{\mathsf{bce}}\omega_{\mathsf{ad}}+\epsilon_{\mathsf{bce}}\omega
_{\mathsf{ad}})\partial^{\mathsf{e}},
\]

\begin{equation}
\mathsf{A}_{\mathsf{ab}}=\epsilon_{\mathsf{abc}}\partial^{\mathsf{c}},
\label{operators}%
\end{equation}

\[
\mathsf{B}_{\mathsf{a,bc}}=\eta_{\mathsf{ab}}\partial_{\mathsf{c}}%
+\eta_{\mathsf{ac}}\partial_{\mathsf{b}}%
\]
and%

\[
\mathsf{D}_{\mathsf{a,bc}}=\mathsf{A}_{\mathsf{ab}}\partial_{\mathsf{c}%
}+\mathsf{A}_{\mathsf{ac}}\partial_{\mathsf{b}}.
\]

We recall that the usual Barnes-Rivers operators obey the algebra:%

\[
\mathsf{P}_{\mathsf{ab,kl}}^{(i-a)}\mathsf{P}^{(j-b)\ \mathsf{kl}}%
{}_{,\mathsf{cd}}=\delta^{ij}\delta^{ab}\mathsf{P}_{\mathsf{ab,cd}}^{(j-b)},
\]

\begin{equation}
\mathsf{P}_{\mathsf{ab,kl}}^{(i-ab)}\mathsf{P}^{(j-cd)\ \mathsf{kl}}%
{}_{,\mathsf{cd}}=\delta^{ij}\delta^{bc}\mathsf{P}_{\mathsf{ab,cd}}^{(j-a)},
\end{equation}

\[
\mathsf{P}_{\mathsf{ab,kl}}^{(i-a)}\mathsf{P}^{(j-bc)\ \mathsf{kl}}%
{}_{,\mathsf{cd}}=\delta^{ij}\delta^{ab}\mathsf{P}_{\mathsf{ab,cd}}^{(j-ac)},
\]

\[
\mathsf{P}_{\mathsf{ab,kl}}^{(i-ab)}\mathsf{P}^{(j-c)\ \mathsf{kl}}%
{}_{,\mathsf{cd}}=\delta^{ij}\delta^{bc}\mathsf{P}_{\mathsf{ab,cd}}^{(j-ac)}%
\]
and satisfy the tensor identity,%

\begin{equation}
\mathsf{P}_{\mathsf{ab,cd}}^{(2)}+\mathsf{P}_{\mathsf{ab,cd}}^{(1m)}%
+\mathsf{P}_{\mathsf{ab,cd}}^{(0s)}+\mathsf{P}_{\mathsf{ab,cd}}^{(0w)}%
=\frac{1}{2}\left(  \eta_{\mathsf{ac}}\eta_{\mathsf{bd}}+\eta_{\mathsf{ad}%
}\eta_{\mathsf{bc}}\right)  . \label{onze}%
\end{equation}

The new set of spin operators that comes about displays, besides the operators
$\mathsf{S}_{\mathsf{ab,cd}}^{(2a)}$, $\mathsf{R}_{\mathsf{ab,cd}}^{(1a)}$,
$\mathsf{A}_{\mathsf{ab}}$, and $\mathsf{B}_{\mathsf{a,bc}}$\ (already known
from \cite{old article}), one new operator, $\mathsf{D}_{\mathsf{a,bc}}$,
given in (\ref{operators}). These five operators have their own multiplicative
table; we quote below only some of the relevant products amongst them:%

\[
\mathsf{S}_{\mathsf{ab,ef}}^{(2a)}\mathsf{S}^{(2a)}{}^{\mathsf{ef}}%
{}_{,\mathsf{cd}}=-16\square\mathsf{P}_{\mathsf{ab,cd}}^{(2)},
\]

\[
\mathsf{R}_{\mathsf{ab,ef}}^{(1a)}\mathsf{R}^{(1a)}{}^{\mathsf{ef}}%
{}_{,\mathsf{cd}}=-4\square\mathsf{P}_{\mathsf{ab,cd}}^{(1m)},
\]

\[
\mathsf{P}_{\mathsf{ab,ef}}^{(2)}\mathsf{S}^{(2a)}{}^{\mathsf{ef}}%
{}_{,\mathsf{cd}}=\mathsf{S}_{\mathsf{ab,ef}}^{(2a)}\mathsf{P}^{(2)}%
{}^{\mathsf{ef}}{}_{,\mathsf{cd}}=\mathsf{S}_{\mathsf{ab,cd}}^{(2a)},
\]

\begin{equation}
\mathsf{P}_{\mathsf{ab,ef}}^{(1m)}\mathsf{R}^{(1a)}{}^{\mathsf{ef}}%
{}_{,\mathsf{cd}}=\mathsf{R}_{\mathsf{ab,ef}}^{(1a)}\mathsf{P}^{(1m)}%
{}^{\mathsf{ef}}{}_{,\mathsf{cd}}=\mathsf{R}_{\mathsf{ab,cd}}^{(1m)},
\end{equation}

\[
\mathsf{A}_{\mathsf{ae}}\mathsf{A}^{\mathsf{e}}{}_{\mathsf{b}}=-\square
\theta_{\mathsf{ab}},
\]

\[
\mathsf{B}_{\mathsf{a,ef}}\mathsf{B}_{\mathsf{c,}}{}^{\mathsf{ef}}%
=2\square(\theta_{\mathsf{ac}}+2\omega_{\mathsf{ac}}),
\]

\[
\mathsf{B}_{\mathsf{e,ab}}\mathsf{B}^{\mathsf{e}}{}_{,\mathsf{cd}}%
=2\square(\mathsf{P}_{\mathsf{ab,cd}}^{(1m)}+2\mathsf{P}_{\mathsf{ab,cd}%
}^{(0w)}),
\]

\[
\mathsf{D}_{\mathsf{a,ef}}\mathsf{D}_{\mathsf{c,}}{}^{\mathsf{ef}}%
=2\square^{2}\theta_{\mathsf{ac}}%
\]
and%

\[
\mathsf{D}_{\mathsf{e,ab}}\mathsf{D}^{\mathsf{e}}{}_{,\mathsf{cd}}%
=2\square^{2}\mathsf{P}_{\mathsf{ab,cd}}^{(1m)}.
\]

Thus, the wave operator acquires the form:%

\begin{equation}
M=\left(
\begin{array}
[c]{ccc}%
yy_{\mathsf{ab,cd}} & yy_{\mathsf{ab,c}} & yh_{\mathsf{ab,cd}}\\
yy_{\mathsf{a,cd}} & yy_{\mathsf{a,c}} & yh_{\mathsf{a,cd}}\\
hy_{\mathsf{ab,cd}} & hy_{\mathsf{ab,c}} & hh_{\mathsf{ab,cd}}%
\end{array}
\right)  ,
\end{equation}
where%

\begin{align*}
yy_{\mathsf{ab,cd}}  &  =(2\mathsf{a}_{1}-2\mathsf{a}_{3}\square
)\mathsf{P}_{\mathsf{ab,cd}}^{(2)}+(2\mathsf{a}_{1}-\mathsf{a}_{3}\square
-2\xi\square)\mathsf{P}_{\mathsf{ab,cd}}^{(1m)}-(2\mathsf{a}_{1}%
+2\mathsf{a}_{3}\square)\mathsf{P}_{\mathsf{ab,cd}}^{(0s)}\\
&  -4\xi\square\mathsf{P}_{\mathsf{ab,cd}}^{(0w)}-2\sqrt{2}\mathsf{a}%
_{1}(\mathsf{P}_{\mathsf{ab,cd}}^{(0sw)}+\mathsf{P}_{\mathsf{ab,cd}}%
^{(0ws)})+\frac{\mathsf{a}_{4}}{2}(\mathsf{S}_{\mathsf{ab,cd}}^{(2a)}%
+\mathsf{R}_{\mathsf{ab,cd}}^{(1a)}),
\end{align*}

\[
yy_{\mathsf{ab,c}}=\mathsf{a}_{4}\mathsf{B}_{\mathsf{c,ab}}+(2\xi
-\mathsf{a}_{3})\mathsf{D}_{\mathsf{c,ab}},
\]

\[
yh_{\mathsf{ab,cd}}=\frac{k\square}{2}\mathsf{a}_{4}(\mathsf{P}%
_{\mathsf{ab,cd}}^{(2)}-\mathsf{P}_{\mathsf{ab,cd}}^{(0s)})+\frac{k}%
{4}\mathsf{a}_{1}(\mathsf{S}_{\mathsf{ab,cd}}^{(2a)}+\mathsf{R}%
_{\mathsf{ab,cd}}^{(1a)}),
\]

\[
yy_{\mathsf{a,cd}}=-\mathsf{a}_{4}\mathsf{B}_{\mathsf{a,bc}}+(2\xi
-\mathsf{a}_{3})\mathsf{D}_{\mathsf{a,bc}},
\]

\begin{equation}
yy_{\mathsf{a,c}}=-(4\mathsf{a}_{1}+2\mathsf{a}_{3}\square+4\xi\square
)\theta_{\mathsf{a,c}}-(4\mathsf{a}_{1}+32\mathsf{a}_{2}\square+12\mathsf{a}%
_{3}\square)\omega_{\mathsf{a,c}}+2\mathsf{a}_{4}\mathsf{A}_{\mathsf{a,c}},
\end{equation}

\[
yh_{\mathsf{a,cd}}=-\frac{k}{2}\mathsf{a}_{1}\mathsf{B}_{\mathsf{a,bc}%
}+k\mathsf{a}_{1}(\theta_{\mathsf{bc}}+\omega_{\mathsf{bc}})\partial
_{\mathsf{a}},
\]

\[
hy_{\mathsf{ab,cd}}=\frac{k\square}{2}\mathsf{a}_{4}(\mathsf{P}%
_{\mathsf{ab,cd}}^{(2)}-\mathsf{P}_{\mathsf{ab,cd}}^{(0s)})+\frac{k}%
{4}\mathsf{a}_{1}(\mathsf{S}_{\mathsf{ab,cd}}^{(2a)}+\mathsf{R}%
_{\mathsf{ab,cd}}^{(1a)}),
\]

\[
hy_{\mathsf{ab,c}}=\frac{k}{2}\mathsf{a}_{1}\mathsf{B}_{\mathsf{a,bc}%
}-k\mathsf{a}_{1}(\theta_{\mathsf{bc}}+\omega_{\mathsf{bc}})\partial
_{\mathsf{a}}%
\]
and%

\[
hh_{\mathsf{ab,cd}}=-\lambda\square\left(  \mathsf{P}_{\mathsf{ab,cd}}%
^{(1m)}+\mathsf{P}_{\mathsf{ab,cd}}^{(0s)}+\frac{1}{2}\mathsf{P}%
_{\mathsf{ab,cd}}^{(0w)}-\frac{\sqrt{2}}{2}(\mathsf{P}_{\mathsf{ab,cd}%
}^{(0sw)}+\mathsf{P}_{\mathsf{ab,cd}}^{(0ws)})\right)  .
\]

In order to calculate the propagators of the theory,%

\begin{equation}
\left\langle 0\right\vert T[F(x)F(y)]\left\vert 0\right\rangle =iM^{-1}%
\delta^{(3)}(x-y), \label{dezoito}%
\end{equation}
we need to calculate the inverse matrix, $M^{-1}$, of the wave operator, but
here we find a problem: the matrix element $hh_{\mathsf{ab,cd}}$ has not a
term in $\mathsf{P}_{\mathsf{ab,cd}}^{(2)}$, and we cannot find the inverse
element of this fundamental term (to compute the inverse we need to close the
relation given in eq.(\ref{onze}), that does not occur).

We can see in this manner that a completely invertible theory, when decomposed
in terms of one gauge field and its torsion tensor components, looses this
property when we focus in the version where we do not adopt the torsion as the
fundamental field, but rather work with the gauge field associated to Lorentz
local transformation that incorporates the torsion information (in a
Einstein-Cartan theory $\omega_{\mathsf{abc}}=\gamma_{\mathsf{abc}%
}-K_{\mathsf{abc}}$, where $\gamma_{\mathsf{abc}}$ is the "pure Riemannian",
without torsion, part and $K_{\mathsf{abc}}$\ is the contortion term). The
missing spin-2 term of the gravitational gauge field is incorporated into the
"Riemannian part" of the spin connection gauge field.

\section{Introducing the Torsion Terms}

In order to try to obtain a pure gauge theory of planar gravitation, and yet
understand the role\ of torsion in it, we change our study to the following action:%

\begin{equation}
\mathcal{S=}\int d^{3}x\ e(\mathsf{a}_{1}\mathcal{R+}\mathsf{a}_{2}%
\mathcal{T}_{\alpha\beta\gamma}\mathcal{T}^{\alpha\beta\gamma}+\mathsf{a}%
_{3}\mathcal{T}_{\alpha\beta\gamma}\mathcal{T}^{\beta\gamma\alpha}%
+\mathsf{a}_{4}\mathcal{T}_{\alpha\beta}{}^{\beta}\mathcal{T}^{\alpha}%
{}_{\gamma}{}^{\gamma}+\mathsf{a}_{5}\mathcal{L}_{CS}),\label{treze}%
\end{equation}
where we explicitly introduce torsion terms in the action, with $\mathcal{L}%
_{CS}$\ being the usual Chern-Simons term given in eq. (\ref{quatorze}).
$\mathsf{a}_{1}$, $\mathsf{a}_{2}$, $\mathsf{a}_{3}$\ and $\mathsf{a}_{4}$ are
free coefficients, whereas $\mathsf{a}_{5}$ is the Chern-Simons parameter. See
reference \cite{artcomVitor} for these specific torsion terms. From now on,
all our calculations and results refer to the action (\ref{treze}). In our
final section, we shall make a comment on the possibility of introducing a
term linear in the torsion \cite{zanelli}.

By means of equations (\ref{nove}), (\ref{doze}) and (\ref{sete}), but the
decompositions (\ref{quinze}), (\ref{dezesseis}) and (\ref{dezessete}) with
the following weak expansion:%

\begin{equation}
e_{\alpha}{}^{\mathsf{a}}=\delta_{\alpha}{}^{\mathsf{a}}+\frac{k}{2}H_{\alpha
}{}^{\mathsf{a}}\left(  \Rightarrow g_{\alpha\beta}=\eta_{\mathsf{\alpha\beta
}}+kh_{\alpha\beta},\ \ h_{\alpha\beta}=\frac{1}{2}(H_{\alpha\beta}%
+H_{\beta\alpha})\right)  .
\end{equation}

With the new decomposition,%

\begin{equation}
H_{\mathsf{ab}}=h_{\mathsf{ab}}+\mathcal{H}_{\mathsf{ab}}\ ,\ h_{\mathsf{ab}%
}=H_{(\mathsf{ab)}}\ \ e\ \ \mathcal{H}_{\mathsf{ab}}=H_{[\mathsf{ab]}}%
\end{equation}
and%

\begin{equation}
\mathcal{H}_{\mathsf{ab}}=\epsilon_{\mathsf{abc}}h^{\mathsf{c}}\Rightarrow
h_{\mathsf{a}}=\frac{1}{2}\epsilon_{\mathsf{abc}}\mathcal{H}^{\mathsf{bc}}.
\end{equation}

We can rewrite the action (\ref{treze}), introducing the gauge-fixing terms%

\begin{equation}
\mathcal{L}_{GF-diff}=\lambda F_{\mathsf{a}}F^{\mathsf{a}}\ ,\ F_{\mathsf{a}%
}=k\partial^{\mathsf{b}}\left(  H_{\mathsf{ba}}-\frac{1}{2}\eta_{\mathsf{ba}%
}H_{\mathsf{c}}{}^{\mathsf{c}}\right)  ,
\end{equation}

in the linearized form:%

\begin{equation}
\mathcal{S}\mathcal{=}\int d^{3}x\frac{1}{2}\Phi^{T}M\Phi\ ,\ \Phi=\left(
\begin{array}
[c]{c}%
h^{\mathsf{cd}}\\
h^{\mathsf{c}}\\
y^{\mathsf{cd}}\\
y^{\mathsf{c}}%
\end{array}
\right)  .
\end{equation}

As before, we express the wave operator, $M$, in terms of the extended
spin-projection operator formalism. In addition to the operators listed above,
there appear two new operators:%

\begin{equation}
\theta_{\mathsf{ab}}\partial_{\mathsf{c}}\text{ \ \ \ \ \ and \ \ \ \ }%
\omega_{\mathsf{ab}}\partial_{\mathsf{c}},
\end{equation}
which, together with the old ones, completely close the algebra.

This yields the form below for the wave operator:%

\begin{equation}
M=\left(
\begin{array}
[c]{cccc}%
hh_{\mathsf{ab,cd}} & hh_{\mathsf{ab,c}} & hy_{\mathsf{ab,cd}} &
hy_{\mathsf{ab,c}}\\
hh_{\mathsf{a,cd}} & hh_{\mathsf{a,c}} & hy_{\mathsf{a,cd}} & hy_{\mathsf{a,c}%
}\\
yh_{\mathsf{ab,cd}} & yh_{\mathsf{ab,c}} & yy_{\mathsf{ab,cd}} &
yy_{\mathsf{ab,c}}\\
yh_{\mathsf{a,cd}} & yh_{\mathsf{a,c}} & yy_{\mathsf{a,cd}} & yy_{\mathsf{a,c}%
}%
\end{array}
\right)  ,
\end{equation}
where%

\begin{align*}
hh_{\mathsf{ab,cd}}  &  =\frac{k^{2}}{2}\square(\mathsf{a}_{3}-2\mathsf{a}%
_{2})\mathsf{P}_{\mathsf{ab,cd}}^{(2)}+\frac{k^{2}}{4}\square(\mathsf{a}%
_{3}-2\mathsf{a}_{2}-\mathsf{a}_{4}-4\lambda)\mathsf{P}_{\mathsf{ab,cd}%
}^{(1m)}\\
&  +\frac{k^{2}}{2}\square(\mathsf{a}_{3}-2\mathsf{a}_{2}-2\mathsf{a}%
_{4}-2\lambda)\mathsf{P}_{\mathsf{ab,cd}}^{(0s)}-(\frac{k^{2}}{2}%
\square\lambda)\mathsf{P}_{\mathsf{ab,cd}}^{(0w)}\\
&  -(\frac{\sqrt{2}}{2}k^{2}\square\lambda)(\mathsf{P}_{\mathsf{ab,cd}%
}^{(0sw)}+\mathsf{P}_{\mathsf{ab,cd}}^{(0ws)})-\frac{\mathsf{k}^{2}}%
{2}\mathsf{a}_{5}(\mathsf{S}_{\mathsf{ab,cd}}^{(2a)}+\mathsf{R}%
_{\mathsf{ab,cd}}^{(1a)}),
\end{align*}

\[
hh_{\mathsf{ab,c}}=-(\frac{k^{2}}{2}\mathsf{a}_{5})\mathsf{B}_{\mathsf{c,ab}%
}+\frac{k^{2}}{4}(\mathsf{a}_{3}-2\mathsf{a}_{2}-\mathsf{a}_{4}+4\lambda
)\mathsf{D}_{\mathsf{c,ab}},
\]

\begin{align}
hy_{\mathsf{ab,cd}}  &  =\frac{k}{2}(\square\mathsf{a}_{6}-2\mathsf{a}%
_{5})\mathsf{P}_{\mathsf{ab,cd}}^{(2)}-(k\mathsf{a}_{5})\mathsf{P}%
_{\mathsf{ab,cd}}^{(1m)}-\frac{k}{2}(\square\mathsf{a}_{6}+2\mathsf{a}%
_{5})\mathsf{P}_{\mathsf{ab,cd}}^{(0s)}\\
&  -(k\mathsf{a}_{5})\mathsf{P}_{\mathsf{ab,cd}}^{(0w)}+\frac{k}{4}%
(\mathsf{a}_{1}+2\mathsf{a}_{2}-2\mathsf{a}_{3})(\mathsf{S}_{\mathsf{ab,cd}%
}^{(2a)}+\mathsf{R}_{\mathsf{ab,cd}}^{(1a)}),\nonumber
\end{align}

\[
hy_{\mathsf{ab,c}}=\frac{k}{2}(\mathsf{a}_{1}-2\mathsf{a}_{2}-2\mathsf{a}%
_{4})\mathsf{B}_{\mathsf{c,ab}}-k(\mathsf{a}_{1}-2\mathsf{a}_{2}%
-2\mathsf{a}_{4})(\theta_{\mathsf{ab}}+\omega_{\mathsf{ab}})\partial
_{\mathsf{c}},
\]

\[
hh_{\mathsf{a,cd}}=(\frac{k^{2}}{2}\mathsf{a}_{5})\mathsf{B}_{\mathsf{a,cd}%
}+\frac{k^{2}}{4}(\mathsf{a}_{3}-2\mathsf{a}_{2}-\mathsf{a}_{4}+4\lambda
)\mathsf{D}_{\mathsf{a,cd}},
\]

\[
hh_{\mathsf{a,c}}=\frac{k^{2}}{2}\square(\mathsf{a}_{3}-2\mathsf{a}%
_{2}-\mathsf{a}_{4}-4\lambda)\theta_{\mathsf{a,c}}-(k^{2}\square
)(2\mathsf{a}_{2}+\mathsf{a}_{3})\omega_{\mathsf{a,c}}-(k^{2}\mathsf{a}%
_{5})\mathsf{A}_{\mathsf{a,c}},
\]

\[
hy_{\mathsf{a,cd}}=-\frac{k}{2}(\mathsf{a}_{1}+2\mathsf{a}_{2})\mathsf{B}%
_{\mathsf{a,bc}}+k(\mathsf{a}_{1}-2\mathsf{a}_{2}-2\mathsf{a}_{3}%
)(\theta_{\mathsf{bc}}+\omega_{\mathsf{bc}})\partial_{\mathsf{a}},
\]

\[
hy_{\mathsf{a,c}}=(2k\mathsf{a}_{5})\theta_{\mathsf{a,c}}+k(2\mathsf{a}%
_{5}-\square\mathsf{a}_{6})\omega_{\mathsf{a,c}}+k(\mathsf{a}_{1}%
-2\mathsf{a}_{2}-2\mathsf{a}_{4})\mathsf{A}_{\mathsf{a,c}},
\]

\begin{align*}
yh_{\mathsf{ab,cd}}  &  =\frac{k}{2}(\square\mathsf{a}_{6}-2\mathsf{a}%
_{5})\mathsf{P}_{\mathsf{ab,cd}}^{(2)}-(k\mathsf{a}_{5})\mathsf{P}%
_{\mathsf{ab,cd}}^{(1m)}-\frac{k}{2}(\square\mathsf{a}_{6}+2\mathsf{a}%
_{5})\mathsf{P}_{\mathsf{ab,cd}}^{(0s)}\\
&  -(k\mathsf{a}_{5})\mathsf{P}_{\mathsf{ab,cd}}^{(0w)}+\frac{k}{4}%
(\mathsf{a}_{1}+2\mathsf{a}_{2}-2\mathsf{a}_{3})(\mathsf{S}_{\mathsf{ab,cd}%
}^{(2a)}+\mathsf{R}_{\mathsf{ab,cd}}^{(1a)}),
\end{align*}

\[
yh_{\mathsf{ab,c}}=\frac{k}{2}(\mathsf{a}_{1}+2\mathsf{a}_{2})\mathsf{B}%
_{\mathsf{c,ab}}-k(\mathsf{a}_{1}-2\mathsf{a}_{2}-2\mathsf{a}_{3}%
)(\theta_{\mathsf{ab}}+\omega_{\mathsf{ab}})\partial_{\mathsf{c}},
\]

\begin{align*}
yy_{\mathsf{ab,cd}}  &  =2(\mathsf{a}_{1}+2\mathsf{a}_{2}-\mathsf{a}%
_{3})\mathsf{P}_{\mathsf{ab,cd}}^{(2)}+2(\mathsf{a}_{1}+2\mathsf{a}%
_{2}-\mathsf{a}_{3})\mathsf{P}_{\mathsf{ab,cd}}^{(1m)}\\
&  +2(6\mathsf{a}_{2}+5\mathsf{a}_{3}-\mathsf{a}_{1})\mathsf{P}%
_{\mathsf{ab,cd}}^{(0s)}+4(2\mathsf{a}_{2}+\mathsf{a}_{3})\mathsf{P}%
_{\mathsf{ab,cd}}^{(0w)}\\
&  +2\sqrt{2}(2\mathsf{a}_{2}+3\mathsf{a}_{3}-\mathsf{a}_{1})(\mathsf{P}%
_{\mathsf{ab,cd}}^{(0sw)}+\mathsf{P}_{\mathsf{ab,cd}}^{(0ws)})+(\frac
{\mathsf{a}_{6}}{2})(\mathsf{S}_{\mathsf{ab,cd}}^{(2a)}+\mathsf{R}%
_{\mathsf{ab,cd}}^{(1a)}),
\end{align*}

\[
yy_{\mathsf{ab,c}}=\mathsf{a}_{6}\mathsf{B}_{\mathsf{c,ab}},
\]

\[
yh_{\mathsf{a,cd}}=-\frac{k}{2}(\mathsf{a}_{1}-2\mathsf{a}_{2}-2\mathsf{a}%
_{4})\mathsf{B}_{\mathsf{a,bc}}+k(\mathsf{a}_{1}-2\mathsf{a}_{2}%
-2\mathsf{a}_{4})(\theta_{\mathsf{bc}}+\omega_{\mathsf{bc}})\partial
_{\mathsf{a}},
\]

\[
yh_{\mathsf{a,c}}=(2k\mathsf{a}_{5})\theta_{\mathsf{a,c}}+k(2\mathsf{a}%
_{5}-\square\mathsf{a}_{6})\omega_{\mathsf{a,c}}+k(\mathsf{a}_{1}%
-2\mathsf{a}_{2}-2\mathsf{a}_{4})\mathsf{A}_{\mathsf{a,c}},
\]

\[
yy_{\mathsf{a,cd}}=-\mathsf{a}_{6}\mathsf{B}_{\mathsf{a,cd}}%
\]
and%

\begin{align}
yy_{\mathsf{a,c}}  &  =4(2\mathsf{a}_{2}+2\mathsf{a}_{4}-\mathsf{a}%
_{1}-\mathsf{a}_{3})\theta_{\mathsf{a,c}}+4(2\mathsf{a}_{2}+2\mathsf{a}%
_{4}-\mathsf{a}_{1}-\mathsf{a}_{3})\omega_{\mathsf{a,c}}\nonumber\\
&  +(2\mathsf{a}_{6})\mathsf{A}_{\mathsf{a,c}}.\nonumber
\end{align}

Once all operators have been identified and worked out, we finally come to the
task of computing the inverses. This is what we shall do next.

\section{Propagators and Excitation Modes}

In order to calculate the propagators, eq. (\ref{dezoito}), we use a
straightforward, but lengthy, procedure in terms of which we decompose the
matrix $M$\ into four sectors, namely:%

\begin{equation}
M=\left(
\begin{array}
[c]{cc}%
hh & hy\\
yh & yy
\end{array}
\right)  .
\end{equation}

Thus the inverse matrix $M^{-1}$\ can be written as:%

\begin{equation}
M^{-1}=\left(
\begin{array}
[c]{cc}%
HH & HY\\
YH & YY
\end{array}
\right)  ,
\end{equation}

where its blocks are given by:%

\begin{align}
HH  &  =[hh-hy(yy)^{-1}yh]^{-1}.\nonumber\\
HY  &  =-(hh)^{-1}hyYY.\label{dezenove}\\
YH  &  =-(yy)^{-1}yhHH.\nonumber\\
YY  &  =[yy-yh(hh)^{-1}hy]^{-1}.\nonumber
\end{align}

Once whit the propagators, we must check the tree-level unitarity of the
theory. To this, we have to analyse the residues of the current-current
transition amplitude in momentum space, given by the saturated propagator
after a Fourier transformation. The sources that saturate the propagators can
be expanded in terms of a complete basis in the momentum space as follows:%

\begin{align}
Sources_{\mu\nu}  &  =c%
\acute{}%
_{1}p_{\mu}p_{\nu}+c%
\acute{}%
_{2}p_{\mu}q_{\nu}+c%
\acute{}%
_{3}p_{\mu}\varepsilon_{\nu}+c%
\acute{}%
_{4}q_{\mu}p_{\nu}+c%
\acute{}%
_{5}q_{\mu}q_{\nu}\\
&  +c%
\acute{}%
_{6}q_{\mu}\varepsilon_{\nu}+c%
\acute{}%
_{7}\varepsilon_{\mu}p_{\nu}+c%
\acute{}%
_{8}\varepsilon_{\mu}q_{\nu}+c%
\acute{}%
_{9}\varepsilon_{\mu}\varepsilon_{\nu},\nonumber
\end{align}
where $p_{\mu}=(p_{0},-\overrightarrow{p})$, $q_{\mu}=(p_{0},\overrightarrow
{p})$\ and $\varepsilon_{\mu}=(0,-\overrightarrow{\varepsilon})$\ are linearly
independent vectors that satisfy the conditions:%

\begin{align}
p_{\mu}p^{\mu}  &  =q_{\mu}q^{\mu}=m^{2}.\nonumber\\
p_{\mu}q^{\mu}  &  =p_{0}^{2}+\overrightarrow{p}^{2}\neq0.\\
p_{\mu}\varepsilon^{\mu}  &  =q_{\mu}\varepsilon^{\mu}=0.\nonumber\\
\varepsilon_{\mu}\varepsilon^{\mu}  &  =-1.\nonumber
\end{align}

These conditions and the symmetry requirements of the theory split the
sources, $S_{\mu\nu}$, in a symmetric and an antisymmetric part:%

\begin{align}
S_{S\mu\nu}  &  =S_{(\mu\nu)}=c_{1}p_{\mu}p_{\nu}+c_{2}(p_{\mu}q_{\nu}+q_{\mu
}p_{\nu})+c_{3}(p_{\mu}\varepsilon_{\nu}+\varepsilon_{\mu}p_{\nu})\\
&  +c_{4}q_{\mu}q_{\nu}+c_{5}(q_{\mu}\varepsilon_{\nu}+\varepsilon_{\mu}%
q_{\nu})+c_{6}\varepsilon_{\mu}\varepsilon_{\nu}\nonumber
\end{align}
and%

\begin{align}
A_{S\mu\nu}  &  =S_{[\mu\nu]}=d_{1}(p_{\mu}q_{\nu}-q_{\mu}p_{\nu}%
)+d_{2}(p_{\mu}\varepsilon_{\nu}-\varepsilon_{\mu}p_{\nu})\\
&  +d_{3}(q_{\mu}\varepsilon_{\nu}-\varepsilon_{\mu}q_{\nu}),\nonumber
\end{align}
where $c_{1}=c%
\acute{}%
_{1},$ $c_{2}=\frac{c%
\acute{}%
_{2}+c%
\acute{}%
_{4}}{2},$ $c_{3}=\frac{c%
\acute{}%
_{3}+c%
\acute{}%
_{7}}{2},$ $c_{4}=c%
\acute{}%
_{5},$ $c_{5}=\frac{c%
\acute{}%
_{6}+c%
\acute{}%
_{8}}{2},$ $c_{6}=c%
\acute{}%
_{9}$ $d_{1}=\frac{c%
\acute{}%
_{2}-c%
\acute{}%
_{4}}{2},$ $d_{2}=\frac{c%
\acute{}%
_{3}-c%
\acute{}%
_{7}}{2},$ and $d_{3}=\frac{c%
\acute{}%
_{6}-c%
\acute{}%
_{8}}{2}$.

The currente-current transition amplitude is written as:%

\begin{align}
\mathcal{A}  &  =\left(
\begin{array}
[c]{cc}%
\tau^{\ast} & \rho^{\ast}%
\end{array}
\right)  \left(
\begin{array}
[c]{cc}%
HH & HY\\
YH & YY
\end{array}
\right)  \left(
\begin{array}
[c]{c}%
\tau\\
\rho
\end{array}
\right)  \Rightarrow\\
\mathcal{A}  &  =\tau^{\ast}HH\tau+\tau^{\ast}HY\rho+\rho^{\ast}YH\tau
+\rho^{\ast}YY\rho,\nonumber
\end{align}
where $\tau$\ is the source to the $h$\ fields and $\rho$\ the source to the
$y$\ fields.

$\mathcal{A}$ can then be cast into the form below:%

\begin{align}
\mathcal{A}  &  =t^{\mathsf{ab}^{\ast}}HH_{\mathsf{ab,cd}}t^{\mathsf{cd}%
}+t^{\mathsf{ab}^{\ast}}HH_{\mathsf{ab,c}}t^{\mathsf{c}}+t^{\mathsf{a}^{\ast}%
}HH_{\mathsf{a,cd}}t^{\mathsf{cd}}+t^{\mathsf{a}^{\ast}}HH_{\mathsf{a,c}%
}t^{\mathsf{c}}\nonumber\\
&  +t^{\mathsf{ab}^{\ast}}HY_{\mathsf{ab,cd}}r^{\mathsf{cd}}+t^{\mathsf{ab}%
^{\ast}}HY_{\mathsf{ab,c}}r^{\mathsf{c}}+t^{\mathsf{a}^{\ast}}%
HY_{\mathsf{a,cd}}r^{\mathsf{cd}}+t^{\mathsf{a}^{\ast}}HY_{\mathsf{a,c}%
}r^{\mathsf{c}}\\
&  +r^{\mathsf{ab}^{\ast}}YH_{\mathsf{ab,cd}}t^{\mathsf{cd}}+r^{\mathsf{ab}%
^{\ast}}YH_{\mathsf{ab,c}}t^{\mathsf{c}}+r^{\mathsf{a}^{\ast}}%
YH_{\mathsf{a,cd}}t^{\mathsf{cd}}+r^{\mathsf{a}^{\ast}}YH_{\mathsf{a,c}%
}t^{\mathsf{c}}\nonumber\\
&  +r^{\mathsf{ab}^{\ast}}YY_{\mathsf{ab,cd}}r^{\mathsf{cd}}+r^{\mathsf{ab}%
^{\ast}}YY_{\mathsf{ab,c}}r^{\mathsf{c}}+r^{\mathsf{a}^{\ast}}%
YY_{\mathsf{a,cd}}r^{\mathsf{cd}}+r^{\mathsf{a}^{\ast}}YY_{\mathsf{a,c}%
}r^{\mathsf{c}},\nonumber
\end{align}
where $t^{\mathsf{cd}}=\tau^{(\mathsf{cd)}}$,\ $t^{\mathsf{c}}=\frac{1}%
{2}\epsilon^{\mathsf{cde}}T_{\mathsf{de}}$ with $T_{\mathsf{de}}=\tau
_{\lbrack\mathsf{de}]}$ and $r^{\mathsf{cd}}=\rho^{(\mathsf{cd)}}%
$,\ $r^{\mathsf{c}}=\frac{1}{2}\epsilon^{\mathsf{cde}}R_{\mathsf{de}}$ with
$R_{\mathsf{de}}=\rho_{\lbrack\mathsf{de}]}$.

Due to the source constraints, $p_{\mathsf{c}}t^{\mathsf{cd}}=0$,
$p_{\mathsf{c}}T^{\mathsf{cd}}=0$, $p_{\mathsf{c}}r^{\mathsf{cd}}=0$ and
$p_{\mathsf{c}}R^{\mathsf{cd}}=0$, only the projectors $\mathsf{P}%
_{\mathsf{ab,cd}}^{(2)}$, $\mathsf{P}_{\mathsf{ab,cd}}^{(0s)}$, $\mathsf{S}%
_{\mathsf{ab,cd}}^{(2a)}$, $\theta_{\mathsf{ab}}\partial_{\mathsf{c}}$and
$\omega_{\mathsf{a,c}}$,give non-vanishing contributions to the amplitude.

For a massless pole, or for a massive pole in the rest frame (where $p_{\mu
}=(m,0)$, $q_{\mu}=(m,0)$\ and $\varepsilon_{\mu}=(0,-\overrightarrow
{\varepsilon})$), only the projectors $\mathsf{P}_{\mathsf{ab,cd}}^{(2)}$ and
$\mathsf{P}_{\mathsf{ab,cd}}^{(0s)}$\ survive and contribute.

With the restrictions above, the amplitude reads:%

\begin{align}
\mathcal{A}  &  =<H2H2_{(2)}>t^{\mathsf{ab}^{\ast}}\mathsf{P}_{\mathsf{ab,cd}%
}^{(2)}t^{\mathsf{cd}}+<H2H2_{(0s)}>t^{\mathsf{ab}^{\ast}}\mathsf{P}%
_{\mathsf{ab,cd}}^{(0s)}t^{\mathsf{cd}}\nonumber\\
+  &  <H2Y2_{(2)}>t^{\mathsf{ab}^{\ast}}\mathsf{P}_{\mathsf{ab,cd}}%
^{(2)}r^{\mathsf{cd}}+<H2Y2_{(0s)}>t^{\mathsf{ab}^{\ast}}\mathsf{P}%
_{\mathsf{ab,cd}}^{(0s)}r^{\mathsf{cd}}\\
+  &  <Y2H2_{(2)}>r^{\mathsf{ab}^{\ast}}\mathsf{P}_{\mathsf{ab,cd}}%
^{(2)}t^{\mathsf{cd}}+<Y2H2_{(0s)}>r^{\mathsf{ab}^{\ast}}\mathsf{P}%
_{\mathsf{ab,cd}}^{(0s)}t^{\mathsf{cd}}\nonumber\\
+  &  <Y2Y2_{(2)}>r^{\mathsf{ab}^{\ast}}\mathsf{P}_{\mathsf{ab,cd}}%
^{(2)}r^{\mathsf{cd}}+<Y2Y2_{(0s)}>r^{\mathsf{ab}^{\ast}}\mathsf{P}%
_{\mathsf{ab,cd}}^{(0s)}r^{\mathsf{cd}},\nonumber
\end{align}
where $<H2H2_{(2)}>$ is the symmetric rank-2 ($H2$ in $H2H2_{(2)}$)
gravitational field propagator associated to the operator $\mathsf{P}%
_{\mathsf{ab,cd}}^{(2)}$ ($_{(2)}$ in $H2H2_{(2)}$). The other coefficients
have analogous meaning. Explicitly writing the sources, we get:%

\begin{align}
\mathcal{A}  &  =\frac{1}{2}(<H2H2_{(2)}>+<H2H2_{(0s)}>)\left\vert
c_{6t}\right\vert ^{2}\nonumber\\
+\frac{1}{2}(  &  <H2Y2_{(2)}>+<H2Y2_{(0s)}>)c_{6t}^{\ast}c_{6r}%
\label{vinte}\\
+\frac{1}{2}(  &  <Y2H2_{(2)}>+<Y2H2_{(0s)}>)c_{6r}^{\ast}c_{6t}\nonumber\\
+\frac{1}{2}(  &  <Y2Y2_{(2)}>+<Y2Y2_{(0s)}>)\left\vert c_{6r}\right\vert
^{2}\nonumber
\end{align}
where $t$ and $r$ in the $c$ mean the source associated to the particular term.

We must now replace the results obtained by the procedure described in
(\ref{dezenove}) into (\ref{vinte}). Before, explicitly we put our results,
the following comments should be done:

\begin{enumerate}
\item With the whole set of action parameters, $\mathsf{a}_{1}$,
$\mathsf{a}_{2}$, $\mathsf{a}_{3}$, $\mathsf{a}_{4}$\ and $\mathsf{a}_{5}$
plus $\lambda$, different from zero, our computational algebraic facilities
failed in attaining a result due the extension of the resulting expressions.

\item Considering the Chern-Simons term, $\mathsf{a}_{5}$,we obtained the
following behaviour in the denominator of the propagator:

\begin{itemize}
\item With $\mathsf{a}_{1}=0$, we have terms proportional to $p^{22}$.

\item The lowest power, $p^{6}$, occurs with $\mathsf{a}_{1}=\mathsf{a}%
_{2}=\mathsf{a}_{4}=0$, only $\mathsf{a}_{3}$\ and $\mathsf{a}_{5}$\ are considered.

\item With $\mathsf{a}_{3}=0$, we do not have\ an invertible case.
\end{itemize}

\item Without the Chern-Simons term, $\mathsf{a}_{5}=0$, we obtain, in all
invertible cases, a power $p^{2}$. This is not a straightforward result; we
may justify it by pointing out that Chern-Simons contributes a term quadratic
in the spin connection with a space-time derivative, whereas the scalar
curvature contributes a term that mixes $H$ with $\omega$. Setting
$\mathsf{a}_{5}$\ to zero, we suppres $\omega-\omega$\ terms with a
derivative, and so we unavoidably reduces the powers of the momentum.
\end{enumerate}

We then consider in (\ref{vinte}) only the cases with $\mathsf{a}_{5}=0$.

The least invertible case occurs by considering only $\mathsf{a}_{3}$
different from zero in the action. In this case, the relevant propagators read:%

\begin{align}
H2H2_{(2)}  &  =\frac{2}{3k^{2}p^{2}\mathsf{a}_{3}}i.\nonumber\\
H2H2_{(0s)}  &  =-\frac{2}{k^{2}p^{2}\mathsf{a}_{3}}i.\nonumber\\
H2Y2_{(2)}  &  =H2Y2_{(0s)}=Y2H2_{(2)}=Y2H2_{(0s)}=0.\\
Y2Y2_{(2)}  &  =\frac{1}{6\mathsf{a}_{3}}i.\nonumber\\
Y2Y2_{(0s)}  &  =0\nonumber
\end{align}
and the saturated amplitude is as given below,%

\begin{equation}
\mathcal{A}=\left(  -\frac{2}{3k^{2}p^{2}\mathsf{a}_{3}}\left\vert
c_{6}\right\vert _{tt}^{2}+\frac{1}{12\mathsf{a}_{3}}\left\vert c_{6}%
\right\vert _{rr}^{2}\right)  i.
\end{equation}

We notice in this expression that the massless pole comes from the $h$-block
and has contributions from the spin-0 and the spin-2 sectors.

Then, by calculating the imaginary part of the residue of the amplitude at the
massless pole, we get:%

\begin{equation}
\operatorname{Im}(res\mathcal{A})=\operatorname{Im}\left(  \lim_{p^{2}%
\rightarrow0}[p^{2}\mathcal{A}]\right)  =-\frac{2\left\vert c_{6}\right\vert
_{tt}^{2}}{3k^{2}\mathsf{a}_{3}}.
\end{equation}

From the requirement of having positive-definite residue at the pole, we must
have $\mathsf{a}_{3}<0$.

Considering now the addition of the scalar of curvature term $\mathsf{a}_{1}$,
we get:%

\begin{align}
H2H2_{(2)}  &  =\frac{2(\mathsf{a}_{3}-\mathsf{a}_{1})}{k^{2}p^{2}%
\mathsf{(}3\mathsf{a}_{3}^{2}+\mathsf{a}_{1}^{2}-3\mathsf{a}_{3}\mathsf{a}%
_{1})}i\nonumber\\
H2H2_{(0s)}  &  =-\frac{2(\mathsf{a}_{3}+\mathsf{a}_{1})}{k^{2}p^{2}%
\mathsf{(\mathsf{a}_{3}^{2}-\mathsf{a}_{1}^{2}+\mathsf{a}_{3}\mathsf{a}_{1})}%
}i\nonumber\\
H2Y2_{(2)}  &  =H2Y2_{(0s)}=Y2H2_{(2)}=Y2H2_{(0s)}=0\\
Y2Y2_{(2)}  &  =\frac{\mathsf{a}_{3}}{2(3\mathsf{a}_{3}^{2}+\mathsf{\mathsf{a}%
_{1}^{2}}-3\mathsf{\mathsf{a}_{3}\mathsf{a}_{1}})}i\nonumber\\
Y2Y2_{(0s)}  &  =0\nonumber
\end{align}
and the amplitude becomes:%

\begin{equation}
\mathcal{A}=\left(  -\frac{2}{k^{2}p^{2}}\times\frac{\mathsf{\mathsf{a}%
_{3}^{3}}}{3\mathsf{a}_{3}^{4}-5\mathsf{\mathsf{a}_{3}^{2}\mathsf{a}_{1}%
^{2}+4\mathsf{a}_{3}a}_{1}^{3}-\mathsf{a}_{1}^{4}}\left\vert c_{6}\right\vert
_{tt}^{2}+\frac{\mathsf{a}_{3}}{2(3\mathsf{a}_{3}^{2}+\mathsf{\mathsf{a}%
_{1}^{2}}-3\mathsf{\mathsf{a}_{3}\mathsf{a}_{1}})}\left\vert c_{6}\right\vert
_{rr}^{2}\right)  i.\nonumber
\end{equation}

We can see that the structure of the amplitude is not changed, with the pole
having contributions from the same spin sectors. The parameters relations now reads:%

\begin{equation}
\operatorname{Im}(res\mathcal{A})=\operatorname{Im}\left(  \lim_{p^{2}%
\rightarrow0}[p^{2}\mathcal{A}]\right)  =-\frac{2}{k^{2}}\frac
{\mathsf{\mathsf{a}_{3}^{3}}}{3\mathsf{a}_{3}^{4}-5\mathsf{\mathsf{a}_{3}%
^{2}\mathsf{a}_{1}^{2}+4\mathsf{a}_{3}a}_{1}^{3}-\mathsf{a}_{1}^{4}}\left\vert
c_{6}\right\vert _{tt}^{2}. \label{vinteum}%
\end{equation}

The denominator in (\ref{vinteum}) can be written as:%

\begin{equation}
\mathsf{(\mathsf{a}_{3}^{2}+\mathsf{a}_{3}\mathsf{a}_{1}-\mathsf{a}_{1}%
^{2})(3\mathsf{a}_{3}^{2}-3\mathsf{a}_{3}\mathsf{a}_{1}+\mathsf{a}_{1}^{2}).}%
\end{equation}

The binomial $\mathsf{3\mathsf{a}_{3}^{2}-3\mathsf{a}_{3}\mathsf{a}%
_{1}+\mathsf{a}_{1}^{2}}$ has complex roots and is greater than zero.

The requirement of having positive-definite residue at the pole implies (with
$\mathsf{a}_{3}<0$) $\mathsf{\mathsf{a}_{1}^{2}-\mathsf{a}_{3}\mathsf{a}%
_{1}-\mathsf{a}_{3}^{2}<0}$. And the scalar term must obey $\frac{1+\sqrt{5}%
}{2}\mathsf{\mathsf{a}_{3}\approx1.618\mathsf{a}_{3}}<\mathsf{\mathsf{a}_{1}%
<}\frac{1-\sqrt{5}}{2}\mathsf{\mathsf{a}_{3}\approx-0.618\mathsf{a}_{3}}$.

The case where all parameters (with exception to $\mathsf{a}_{5}$) are
different from zero brings only new algebraic corrections to the amplitude,
without changing its structure. The relations among the parameters become very
cumbersome, due to the considerable number of parameters involved, so that
many hypotheses must be done.

\section{Concluding Comments}

In the course of the calculations we report on here, if we complete the action
(\ref{treze}) by adjoining the term $\mathsf{a}_{6}\epsilon^{\mu\nu\lambda
}\mathcal{T}_{\mu\nu}{}^{\mathsf{\mathsf{a}}}e_{\lambda}{}^{\mathsf{b}}%
\eta_{\mathsf{\mathsf{ab}}}=\mathsf{a}_{6}\epsilon^{\mu\nu\lambda}%
\mathcal{T}_{\mu\nu}{}^{\mathsf{\alpha}}e_{\alpha}{}^{\mathsf{a}}e_{\lambda}%
{}^{\mathsf{b}}\eta_{\mathsf{\mathsf{ab}}}=\mathsf{a}_{6}\epsilon^{\mu
\nu\lambda}\mathcal{T}_{\mu\nu\lambda}$ \cite{zanelli}, a\ problem shows up:
though our procedure of introducing the spin operators works, the propagators
could not be found in their generality (with all the six coefficients
$\mathsf{a}_{i}$) even with the help of algebraic computation techniques.
However, we found out that, once any of the $\mathsf{a}_{i}$ are set to zero,
we succeed in reading off the propagators, even if they display higher powers
in the momentum. It is worthwhile to mention here that this linear term in the
torsion combines with the Chern-Simons action to give a rich structure of
poles in the propagators. We do not report these results here because this
investigation is the matter of a forthcoming publication \cite{futurohelay}.
The situation gets better when we discovered that, ruling out the Chern-Simons
term, we get only simple poles in the terms that contribute to the amplitude.
Very surprising was the discovery of the very different role the torsion terms
($\mathsf{a}_{2}$ and $\mathsf{a}_{3}$) play, being $\mathsf{a}_{3}$
fundamental to compute the inverse matrix, which is not the case for
$\mathsf{a}_{2}$.

We see that the physical poles are all massless. It is worthy to note that, in
\cite{old article}, we get only physical mass poles. The unitarity condition
for the physical poles demand that $\mathsf{a}_{3}<0$ and this implies in that
the parameter that governs the scalar curvature must obey the condition
$\frac{1+\sqrt{5}}{2}\mathsf{\mathsf{a}_{3}}<\mathsf{\mathsf{a}_{1}<}%
\frac{1-\sqrt{5}}{2}\mathsf{\mathsf{a}_{3}}$.

\section{Acknowledgments}

The authors acknowledge Prof F. W. Hehl for discussions and suggestions. They
also express their gratitude to CNPq-Brasil for the invaluable financial help.


\section{References}

\begin{thebibliography}{99}                                                                                               %


\bibitem {two}H I Arcos and J G Pereira, \textit{Int. J. Mod. Phys. }D
\textbf{13 }(2004)\textbf{ }2193

\bibitem {old article}J L Boldo, L M de Moraes and J A Helay\"{e}l-Neto,
\textit{Class. Quantum Grav.} \textbf{17} (2000) 813

\bibitem {hehl1}F W Hehl, P von der Heyde, G D Kerlick and J M Nester,
\textit{Rev. Mod. Phys.} \textbf{48} (1976) 3641

\bibitem {hehl2}F W Hehl, J D McCrea, E W Mielke and Y Ne'eman, \textit{Phys.
Rep.} \textbf{258} (1995) 1

E W Mielke and P Baekler, \textit{Phys. Lett. A }\textbf{156} (1991) 399

P Baekler, F W Hehl and E W Mielke, Nonmetricity and Torsion: Facts and
Fancies in Gauge Approaches to Gravity. Rome 1985 Proceedings, General
Relativity A, 277 

\bibitem {novello}M Novello, \textit{Phys. Lett. A} \textbf{59} (1976) 105

\bibitem {sabbata2}V de Sabbata and M Gasperini, \textit{Phys. Lett. A
}\textbf{77} (1980) 300

\bibitem {carrol}S M Caroll and G B Field, \textit{Phys. Rev. D} \textbf{50}
(1994) 3867

\bibitem {hammond}R T Hammond, \textit{Phys. Rev. D} \textbf{52} (1995) 6918

\bibitem {shapiro}A S Belyaev and I L Shapiro, \textit{Phys. Lett. B
}\textbf{425} (1998) 246

\bibitem {sabbata}V de Sabbata and M Gasperini, \textit{Introduction to
Gravitation }(World Scientific) (1985)

I L Shapiro, \textit{Mod. Phys. Lett. A }\textbf{9} (1994) 729

I L Shapiro, \textit{Phys. Rept.} \textbf{357} (2002) 113

\bibitem {zanelli}J Zanelli, (Super-) Gravities beyond Four Dimensions,
\textit{Proceedings of the Summer School Villa de Leyva }(World Scientific) (2003)

J Zanelli, (Super-) Gravidades de Chern-Simons, \textit{Anais da V Escola do
CBPF}, vol. 2, Cursos de P\'{o}s-Gradua\c{c}\~{a}o (CBPF) (2005)

\bibitem {rivers}R J Rivers, \textit{Nuovo Cimento} \textbf{34} (1964) 387

\bibitem {nieuwenhuizen}P van Nieuwenhuizen, \textit{Nucl. Phys. B}
\textbf{60} (1973) 478

F C P Nunes and G O Pires, \textit{Phys. Lett.} B \textbf{301} (1993) 339

\bibitem {artcomVitor}E Sezgin and P van Nieuwenhuizen, \textit{Phys. Rev. D}
\textbf{21} (1980) 3269

\bibitem {futurohelay}L M de Moraes, J A Helay\"{e}l-Neto and V J V\'{a}squez
Otoya, work in progress
\end{thebibliography}
\end{document}